----------------------------------------------------------------------------
\documentstyle{l-aa}                          
\def\ros{{\sl ROSAT }}
\def\etal{{et al. }}
\def\ein{{\sl Einstein }}
\def\grad{$^\circ$}
\def\it{\sl}

\begin{document}

   \thesaurus{08                     
              (09.19.2 G272.2--3.2;  
               08.16.7 PSR 0905-51;  
               13.25.4;              
               02.19.1)              

   \title{G272.2--3.2: A centrally filled and possibly young SNR
            discovered by ROSAT}

   \author{J. Greiner, R. Egger, B. Aschenbach}

   \offprints{J. Greiner}

   \institute{Max-Planck-Institut f\"ur extraterrestrische Physik,
              85740 Garching, Germany}

   \date{Received February 3, 1994; accepted May 6, 1994}

   \maketitle
   \markboth{Greiner, Egger, Aschenbach: G272.2--3.2}{Greiner, Egger,
           Aschenbach: G272.2--3.2}

   \begin{abstract}
We report the discovery of a new SNR in the \ros All-Sky-Survey data.
The SNR has a nearly circular shape with a diameter of about 15 arcmin.
The surface brightness is rather uniform over the remnant. The observed
interstellar absorption suggests a distance of about 1.8$^{+1.4}_{-0.8}$ kpc.

The X-ray emission appears to be of thermal origin with a plasma temperature
of about 10$^7$ K. The Sedov model leads to an age of only 1800 years.
We find no spectral evidence of non-thermal emission in the center of the
remnant. The two most likely explanations for the centrally filled structure
are evaporation of shock heated clouds in the interior of the remnant or
a reverse shock heating the ejecta of the SNR explosion.

   \keywords{supernova remnants -- pulsars -- interstellar matter}
   \end{abstract}

\section{Introduction}

The \ros All-Sky-Survey represents a unique database for the discovery of new
X-ray SNR's. As already reported earlier there exist bright
X-ray SNR's which are not or only marginally visible at radio wavelengths
(Pfeffermann \etal 1991, Aschenbach 1993).
Here, we report details of the X-ray emission of a relatively young,
centrally filled SNR discovered in the \ros All-Sky-Survey data which has
no obvious radio counterpart.

\section{Observational results}

A new bright, extended X-ray source (hereafter called G272.2--3.2) centered on
R.A. = 9$^h$06$^m$40$^s$ and Decl. = --52\grad05'25" (equinox J2000.0)
was discovered in the ROSAT All-Sky-Survey (Greiner \& Egger 1993).
G272.2--3.2 was observed in November 1990 for about 850 sec with the
\ros position sensitive proportional counter (PSPC). The mean PSPC source
count rate of G272.2--3.2 is
(3.9$\pm$0.2)$\times$10$^{-3}$ cts s$^{-1}$ arcmin$^{-2}$ resulting in a
total of 600 counts used in the spectral analysis described below.

Fig. 1 shows the image of G272.2--3.2 with
a resolution of 60 arcsec corresponding to the mean half width of the ROSAT
X-ray telescope point spread function in the survey mode.
The extended emitting region of G272.2--3.2 is nearly circular-shaped
with an X-ray size of about 15.2 arcmin (FWHM), and rather uniform in
brightness with a number of blobs with enhanced emission.

The spectrum of the extended emission is rather hard and considerably absorbed
(no photons are detected below 0.5 keV). Several spectral models have been fit
to the data (Tab. 1) and the results will be discussed
in the next paragraph. Due to statistical limitations (600 cts) it is not
possible to analyze parts of the X-ray emitting region of G272.2--3.2 (as
for instance individual blobs).

On the ESO/SRC  plate R 6712 (2 hr exposure of 1986 Dec. 28)
there is a faint nebulosity near the center of G272.2--3.2 and another, more
extended one about 6 arc\-min west of the center (Fig. 2). Though barely
visible, one recognizes tiny filaments along the perimeter of the X-ray
emitting region, one of which appears to correspond with an X-ray blob
(upper right arrow in Fig. 2). Additionally, the optical background
brightness appears to be enhanced within the contours of the X-ray emitting
region.

A spectroscopic observation (Winkler \etal 1993) of the optical nebulosity
near the center of G272.2--3.2 revealed the ratio of [S II] to H$_\alpha$
emission to be 1.4. This is indicative of shock-heated material. Also, the
[N II] and [O II] emission lines found are typical of SNR's.

No radio counterpart of G272.2--3.2 was found in available radio catalogues.
This may be explained by the fact that radio measurements are hindered by the
vicinity of the radio-loud  Vela-SNR (2\grad~ distance). For instance,
the 408 MHz catalog (Haslam \etal 1982)
shows a strong gradient at the SNR position with a flux density between
50--100 Jy which primarily stems from Vela.
Any point source (radio counterpart of G272.2--3.2) had to be brighter
than this level in order to be detectable.

   \begin{figure}[thbp]
      \caption[\ros image]{Smoothed maximum likelihood image of the
        new supernova remnant G272.2--3.2 taken with the ROSAT PSPC
        during the All-Sky-Survey in November 1990.
        Contours are at likelihood 1, 5, 9, 13, 17, 19, 21.
        Overlayed is a equatorial (equinox 2000.0) coordinate grid.
        }
      \label{survey}
   \end{figure}

\section{Discussion}

\subsection{Morphology}

The great morphological variety of SNR's led to the attempt to divide them
into three classes based on their radio properties (Weiler 1985).
Shell type SNR's typically consist of steep-spectrum radio emission rings.
Plerionic SNR's are diffuse, centrally filled synchrotron nebulae with a
flat radio spectrum.
Composite SNR's consist of a plerionic nebula surrounded by a shell.

The image (Fig. 1) clearly shows that G272.2--3.2
has a nearly circular shape and that the surface brightness is rather smooth
over the entire remnant (except several blobs). The SNR does not appear to be
a simple shell type. The comparison with the SNR's observed with the
\ein satellite (Seward 1990) shows that there is no known similar SNR except
possibly G292.0+1.8 = MSH 11--54. Based on its radio morphology, G292.0+1.8
is classified as a composite remnant (with some uncertainty). At X-ray
energies,
however, the faint, extended emission of G292.0+1.8 appears to be circular
with the bulk
of the emission coming from the central area which is considerably smaller than
the radio extent. In most cases studied so far, the morphology of the emission
pattern at X-rays is similar to the radio morphology.
Thus, it would be interesting to measure the radio morphology
of G272.2--3.2 and to see how it compares to known X-ray SNR's.

   \begin{figure}[htbp]
      \caption[optical image]{Optical appearance of G272.2--3.2 after
         enlargement of the ESO  Schmidt  plate R 6712.
         The cross marks the central position as given in Tab. 1.
         The arrows indicate the three optical filaments examined by
         Winkler \etal (1993).}
      \label{opt}
   \end{figure}

\subsection{Spectral modelling}

Several models were fit to the
data including power law, pure thermal bremsstrahlung and a thermal plasma
model after Raymond-Smith. A thermal equilibrium plasma model gives the
best fit to the \ros PSPC data (Fig. 3) of the integral emission
($\chi^2_{red}$ = 0.76 for 11 degrees of
freedom). The modelling yields a temperature of
about T = 1.4$\times$10$^7$ K, an emission measure of 0.4 cm$^{-6}$pc and
an absorbing column density of N$_H$ = (4.6$\pm$3)$\times$10$^{21}$ cm$^{-2}$.
The observed energy flux of G272.2--3.2 within the \ros energy band
(0.08--2.4 keV) is F$_{abs}$ = 8.0$\times10^{-12}$ erg cm$^{-2}$ s$^{-1}$
and the inferred unabsorbed flux is
F$_{unabs}$ = 3.0$\times10^{-11}$ erg cm$^{-2}$ s$^{-1}$.

   \begin{figure}[thbp]
      \caption[spectrum]{
Results of the spectral analysis: The left panel shows the
\ros PSPC countrate spectrum and the best-fit thermal
bremsstrahlung model (see Tab. 1 for the parameters). The right panel shows
the error ellipse for the temperature versus  N$_H$ of G272.2--3.2.
The contours are at the 1$\sigma$, 2$\sigma$ and 3$\sigma$ confidence level.
      }
      \label{spec}
   \end{figure}

Using the relation of N$_H$[10$^{21}$] $\approx$ 2.2 A$_V$
(Gorenstein 1975) and A$_V$ $\approx$ 1.9 mag/kpc (Allen 1973),
the measured N$_H$ value translates into a distance of 1.1 kpc.
Observations of bright X-ray binaries in or near the galactic plane with
ROSAT and a correlation to the observed optical extinction of the
corresponding counterparts have resulted in N$_H$[10$^{21}$] $\approx$ 1.3
A$_V$
(Predehl 1992). This would give 1.8 kpc for the distance of G272.2--3.2.
Since the latter relation is derived from a larger sample of X-ray sources
with a larger spread in their parameters, the distance of 1.8 kpc will
be adopted in the following discussion. The error estimate based just on the
variation of the fit parameters (see Fig. 3) results in 1.8$^{+1.4}_{-0.8}$
kpc.
At this distance, the SNR luminosity would be in the range
(0.2--3.5)$\times$10$^{34}$ erg/s in the 0.08--2.4 keV band.

Under the assumption that the measured temperature represents the
average temperature behind the shock front of the blast wave,
an expansion velocity of $\approx900$ km/s is derived. The ambient density
of the undisturbed medium would be $n_0=0.2$ cm$^{-3}$ and thus, the total
mass swept up by the blast wave would be about 1 M$_{\odot}$.
The application of the Sedov model (Sedov 1959) for the adiabatic evolution
of a SNR leads to an age of 1.8$\times10^3$ yr and an initial energy of
$E_0\approx2.5\times10^{49}$ erg. This, however, makes it doubtful, whether
the Sedov solution is applicable in such an early stage of the
SNR evolution. The derived initial energy is significantly lower
than the standard values of 10$^{50}$--10$^{51}$ erg from supernova
explosion models.

Since $E_0$ $\propto$ d$^{2.5}$, increasing the distance from the above
1.8$^{+1.4}_{-0.8}$ kpc by a factor of 2--3 would result in a
(2--3)$^{2.5}$ = 6--15 times higher value for the initial energy, i.e.
to values comparable to the standard ones.
If one takes the initial energy to be 10$^{51}$ erg as basic assumption
(as has been done by Pfeffermann \etal (1991) and Kassim \etal (1993) for \ros
observations of other remnants), then the derived distance would be 9 kpc
with a height below the plane of 500 pc. But even at this large distance, the
Sedov age is only 9$\times10^3$ yr (still young).

Thus, one would have to change at least two parameters in order to push
the Sedov solution to reasonable results.
If one takes the lowest temperature allowed by the fit,
which corresponds to a higher absorption (right panel of Fig. 3) and thus
a larger distance of about 3.2 kpc, the derived initial energy is just
above 10$^{50}$ erg. However, since the luminosity is sensitive to the fitted
temperature, these parameters would result in a luminosity of
8.5$\times$10$^{36}$ erg/s which is rather high at the low ambient density.
Thus, the above discussed variation of fit parameters does not lead to
a consistently reasonable result if the Sedov model is used.
Therefore, one is forced to conclude that G272.2--3.2 has not yet reached
the adiabatic stage.

The low swept-up mass indicates that the deceleration of the blast wave may
still be insignificant which in turn supports the idea that the adiabatic
stage may not have been reached, yet. Moreover it is not clear, whether the
measured X-ray temperature is a well defined measure for the expansion
velocity.
There is evidence for some evolutionary young SNR's that
collisional ionization equilibrium has not been reached and lacking
equipartition between electron and ion temperatures may be
important (Nugent \etal 1984). In this case, fitting the
X-ray spectrum with an equilibrium model may lead to wrong results.
Consideration of early phases of SNR evolution and application of
non-equilibrium models  would require X-ray data with
higher spatial and spectral resolution. However, the number of photons
detected is not sufficient to allow such a study with the ROSAT Survey data.
Assuming that the shock wave has not been significantly decelerated,
a velocity of roughly 8000 km/s, like in Cas-A, may also be representative for
G272.2--3.2. This would result in an age of only 400 years, suggesting that
G272.2--3.2 is a historical SNR.

Assuming alternatively that G272.2--3.2 is a plerionic SNR,
the X-ray spectrum is expected to follow a flat power-law, characteristic of a
synchrotron nebula possibly powered by a pulsar.
However, a power-law model does not fit the data well and gives
an untypical photon index of $\alpha=-12$. Therefore, we conclude that a pure
synchrotron origin of the radiation can be ruled out.

Another possibility would be that G272.2--3.2 is a composite type SNR.
Therefore, we tried to fit a power-law model to the spectrum of the
central blob only. The statistics is too poor to either exclude or confirm
the presence of synchrotron emission from the center. Thus, the interpretation
of the centrally filled emission by a composite SNR model cannot be ruled out.
Low statistics also prevent any time variability analysis
and thus the check for a possible central pulsar.

Besides the composite model there are two other possibilities to explain the
centrally filled X-ray structure:
\begin{itemize}
\item The evaporation of dense clumps of material bathed in the hot gas behind
the SN shock may provide the reservoir to increase the density of the SNR core
(McKee \& Ostriker 1977). For instance, a centrally intensity enhancement (as
in SNR's W28 and 3C 400.2) can be adequately modelled under the
assumption that the evaporation time scale is about
10-50 times larger than the age of the corresponding remnant (Long \etal 1991).
\item The blast wave will start decelerating when an amount of
interstellar matter comparable to the initial mass of the SN ejecta has
been swept up. At this point, a reverse shock forms which may
thus reheat the stellar ejecta. This process could lead to a
central luminosity enhancement. A distinction between these scenarios
is presently not possible. Also, mapping of the radio morphology would help in
understanding the emission mechanism of G272.2--3.2. No additional
radio emission would be expected from the interior of the remnant, if cloud
evaporation were responsible for the central X-ray enhancement.
\end{itemize}

\subsection{Possible relation to PSR 0905--51}

It is interesting to note that the pulsar PSR 0905-51 is located
near the north-eastern rim of G272.2--3.2.
The pulsar PSR 0905-51 has not been detected in X-rays; the 3 $\sigma$ upper
limit for its X-ray emission from the survey data
is 4$\times$10$^{-12}$ erg cm$^{-2}$ s$^{-1}$ (0.02 cts/sec)
in the hard band (0.9-2.4 keV) and 2$\times$10$^{-12}$ erg cm$^{-2}$ s$^{-1}$
(0.013 cts/sec) in the soft band (0.08-0.4 keV), assuming 10$^6$ K blackbody
radiation and an absorbing column density of 4.5$\times$10$^{21}$ cm$^{-2}$.

It is tempting to construct a physical association of PSR 0905-51 and
G272.2--3.2. However, the pulsar's spin-down age of 2$\times$10$^6$ years is
considerably larger than the age which we deduce for G272.2--3.2.
Moreover, the present apparent position of PSR 0905-51 would imply an
average transversal pulsar velocity of a few thousand km/s, close to that of
the shock front itself which would be rather unusual. The distance of
G272.2--3.2 of about 1.8$^{+1.4}_{-0.8}$ kpc compares to the pulsar's distance
of d$\approx$2.65 kpc (Taylor \etal 1993). However, due to the large angular
size and proximity of the Gum Nebula, a large fraction of interstellar matter
along the line of sight towards PSR 0905-51 may be highly ionized.
Thus, the estimated pulsar distance based on the dispersion measure,
is rather uncertain, ranging between 1--3.7 kpc (Taylor \etal 1993,
Taylor \& Cordes 1993). Though the distances may be reconciled, we think
that a physical association of PSR 0905-51 and G272.2--3.2 is unlikely.

   \begin{table}
      \caption{Observed and deduced parameters of G272.2--3.2}
            \begin{tabular}{lc}
            \hline
            \noalign{\smallskip}
            \multicolumn{2}{c}{observed parameters}\\
            \noalign{\smallskip}
            source count rate  & 0.7$\pm$0.1 cts/sec \\
            surface brightness &
                    (3.9$\pm$0.2)$\times$10$^{-3}$ cts s$^{-1}$arcmin$^{-2}$\\
            Angular radius  & 7.6 arcmin  \\
            \noalign{\smallskip}
            approx. center of SNR      & $\alpha_{(2000.0)} = 9^h06^m40^s$ \\
                                  & $\delta_{(2000.0)} = -52^{\circ}05'25''$ \\
            \noalign{\medskip}
            \multicolumn{2}{c}{best~fit~values: (thermal~plasma~model)} \\
            \noalign{\smallskip}
            N$_H$ &      (4.6$\pm$3)$\times$10$^{21}$ cm$^{-2}$  \\
            temperature T & (1.4$\pm$0.3)$\times$10$^{7}$ K\\
            unabsorbed energy flux
                       & 3.0$\times10^{-11}$ erg cm$^{-2}$ s$^{-1}$\\
            \noalign{\medskip}
            \multicolumn{2}{c}{deduced parameters} \\
            \noalign{\smallskip}
            Distance D & 1.8$^{+1.4}_{-0.8}$ kpc \\
            Radius R & 4.0 pc \\
            expansion velocity v & 900 km/s \\
            ``Sedov Age'' & 1.8$\times10^{3}$ years \\
            Luminosity L$_x$  & 1.1$\times10^{34}$ erg/s (0.08--2.4 keV) \\
            \noalign{\smallskip}
           \hline
           \end{tabular}
           \label{results}
   \end{table}

\section{Summary}

We have discovered a nearly circular shaped and centrally filled SNR in the
\ros All-Sky-Survey data. The spectrum appears to be of thermal origin.
Depending on the applicability of the Sedov model, the derived age is
between 400 and 4000 yrs. In any case, G272.2--3.2 seems to be a young SNR.

There is no spectral evidence for the emission to be of
non-thermal origin. The morphological structure does not support
a plain shell-type SNR. The most likely explanations for the  observed
morphology are evaporation of shock heated clouds in the interior of the
remnant, or a reverse shock heating the ejecta of the SNR explosion.
A physical association of the radio pulsar PSR 0905-51, located near the
north-eastern rim of G272.2--3.2, is unlikely.

   \acknowledgements
We thank the EXSAS group for their support with the data analysis software and
an unknown referee for valuable comments. This research has made use of the
Simbad database, operated at CDS, Strasbourg, France.
JG was partly supported by the Deutsche Agentur f\"ur
Raumfahrtangelegenheiten (DARA) GmbH under contract number FKZ 50 OR 9201.

\end{document}